\documentclass[12pt]{article}
\usepackage{graphics}
\usepackage{bm}
\usepackage{graphicx}
\usepackage{amsmath}
\usepackage{amssymb}
\usepackage{amstext}
\usepackage{amscd}
\usepackage{amsfonts}
\usepackage{appendix}
\usepackage{color}
\usepackage{wasysym}
\usepackage{latexsym}
\DeclareMathAlphabet{\EuFrak}{U}{euf}{m}{n}
\DeclareMathAlphabet{\EuScript}{U}{eus}{m}{n}
\newcommand{\nd}{\noindent}

\newcommand{\be}{\begin{equation}}
\newcommand{\ee}{\end{equation}}
\newcommand{\ben}{\begin{eqnarray}}
\newcommand{\een}{\end{eqnarray}}

\title{{\bf Reciprocity relations and generalized entropic quantifiers that lack trace-form}}

\author{{A.Plastino$^{1,2}$, A.R.Plastino$^{3}$, M.C.Rocca$^{1,2}$}\\
\small{$^1$Departamento de F\'{\i}sica, Fac. de Ciencias Exactas},\\
\small{Universidad Nacional de La Plata}\\
\small{C.C. 727 (1900) La Plata, Argentina}\\
\small{$^2$IFLP-CCT-CONICET-C.C 727 (1900) La Plata. Argentina}\\
\small{$^3$ CeBio y Secretaria de Investigacion,}\\
\small{Univ. Nac. del Noroeste de la Prov. de Bs. As.,}\\
\small{ UNNOBA and CONICET, R. Saenz Pe\~{n}a 456, Junin, Argentina}}

\date{\today}

\begin{document}

\maketitle

\begin{abstract}

In this effort we show that the Legendre reciprocity relations,   thermodynamics’ essential formal feature, are respected by any entropic functional, even if it is NOT of trace-form nature, as Shannon’s is.
Further, with reference to the MaXent variational process, we encounter important cases, relevant to physical applications currently discussed in the research literature, in which the associated reciprocity relations exhibit anomalies.
We show that these anomalies can be cured by  carefully discriminating between  apparently equivalent entropic forms.

\vskip 2mm


\end{abstract}

%\maketitle

\section{Introduction}

\nd 
\nd Renyi's entropy  $S_R$ is an important quantifier 
in variegated areas of scientific activity. We mentions as examples 
 ecology, quantum information, the Heisenberg
XY spin chain model, theoretical computer science, conformal
field theory, quantum quenching, diffusion processes, etc.  \cite{1,2,3,4,5,6,7,8,9,10,zentro}.   and references therein.  A typical Renyi-feature is the lack of trace form, since  $S_R$ is not of the form

\be S= \int dV f(\xi), \label{trace}\ee with dV the appropriate volume element, $\xi$ a probability density (PD), and $f$ an arbitrary smooth function of the PD. 
Instead, $S_R$ is the logarithm of $S$ above, for $f = [1/(q-1)] \xi^q$, $q$ a real number.
\vskip 3mm

\nd In this work, we  focus attention on thermodynamics' reciprocity relations and re-visit some issues concerning generalized entropies that, we believe,  lack yet full adequate understanding. We focus attention on   the canonical ensemble. Our aims are:

\begin{enumerate}

\item To establish whether general entropies that lack trace form, with Renyi's logarithm replaced by an arbitrary smooth functional $G$ of $f$ in (\ref{trace}), can be successfully described 
  by Jaynes' MaxEnt variational treatment, so that reciprocity relations hold \cite{jaynes}.

\item To analyze anomalies that sometimes arise with regards to the workings of MaxEnt's Lagrange multipliers 
\cite{11}.

\item To assess whether these anomalies can be eliminated.

\end{enumerate}

\section{Background: Legendre transform  and reciprocity relations}

%\subsection{Legendre transform  and reciprocity relations}

For the Lagrange multipliers in the canonical ensemble  we use this notation. 
\begin{itemize}
\item $\lambda_U$ is the energy $U$ multiplier,
\item $\lambda_N$  is the normalization multiplier.
\end{itemize}
In statistical mechanics, these multipliers are always endowed with  meaningful physical information 
\cite{book}.

\nd     Legendre's transform  (LT)  is an operation that converts a
real function  $f_1$ with real variable $x$ into another
$f_2$, of another variable $y$, keeping constant the
information content of $f_1$. The derivative of  $f_1$ becomes
the argument of $f_2$. \be f_2(y)= xy -f_1(x);\,\,\,
y=f_1'(x)\Rightarrow\,\,{\rm reciprocity}.\ee

\nd    LT' {\it reciprocity relations are thermodynamics'
basic formal ingredient} \cite{deslogue}. For two 
functions $S$ and $\mu_J$ one has

 \ben \label{eq.1-13a} S(<A_1>,\ldots, <A_M>) \,=\,\mu_J
  + \sum_{k=1}^M~\mu_k \langle A_k \rangle  , \een with the $A_i$ extensive
  variables and the  $\mu_i$ {\it independent} intensive ones.
 Obviously, the Legendre transform main goal is that of  changing the identity
of our relevant independent  variables. For $\mu_J$ we have

 \be
\label{eq.1-13b}\mu_J(\mu_1,\ldots,\mu_M)= S-
\sum_{k=1}^M~\mu_k\left\langle A_k\right\rangle. \ee Note that for general entropic measures (other than Shannon-Gibbs') $\mu_J=\mu_J(\mu_1,\ldots,\mu_M) $ does not coincide with the normalization Lagrange multiplier. 
The three
operative reciprocity relations become \cite{deslogue}

\be \label{RR-1} \frac{\partial \mu_J}{\partial \mu_{k}}= -
\langle A_k\rangle ~; \hspace{1.cm}  \frac{\partial S }{\partial
\langle A_k \rangle}\,=\,\mu_k  ~ ;\hspace{1.cm}
\frac{\partial S}{\partial \mu_{i}}=\sum_{k}^{M} \mu_{k}
 \frac{\partial \langle A_k \rangle}{\partial \mu_{i}},\ee
the last one being the so-called Euler theorem. 
 In Jaynes' philosophy \cite{jaynes} $S$ is an information amount, to be maximized subject to {\it a priori known} values for the constraints $\langle A_k \rangle$.

%%%%%%%%%%%%%%%%%%%%%%%%%%%%%%%%%%%%%%%%%%%%%ECTION2  %%%%%%%%%%%%%%%%%%%%%%%%%
	\section{General not-trace-form entropies and reciprocity}
	All trace form entropies can be successfully described 
  by Jaynes' MaxEnt variational treatment, so that reciprocity relations hold, as demontrted in 
	\cite{universal}. Let us consider the general lack-of-trace-form instance (see (\ref{trace} in the Introduction). One has, in the language of the Introduction,
		
		\be S= G\left[\int dV f(\xi)\right],   \ee
with $G$ an arbitrary smooth function. 
 Then
\be S'= G' \int dV f'(\xi).\ee 
(Here $S'$ denotes the functional derivative).
Define $F=G'[f'(\xi)] $ and consider the inverse function of $F$, namely,

\be g= F^{-1}.\ee The MaxEnt variational problem ends up being

\be F - \lambda_N -\lambda_U U=0,\ee so that the MaxEnt solution's PD $\xi_{ME}$ is

\be  \xi_{ME}= g(\lambda_N +\lambda_U U), \label{onebis}\ee and the MaxEnt entropy reads

\be  S_{ME}=  G[\int dV f[g(\lambda_N +\lambda_U U)]].\label{twobis}\ee One also has

\be \frac{\partial \langle U \rangle }{\partial \lambda_U }= 
  \int dV U g'(\lambda_N +\lambda_U U) [\frac{\partial \lambda_N}{\partial \lambda_U}+U], \label{threebis}\ee  and

\ben 0= \frac{\partial }{\partial \lambda_U}\int dV \xi =\\= 
   \int dV g'(\lambda_N +\lambda_U U) [\frac{\partial \lambda_N}{\partial \lambda_U}+U]=0,  \label{fourbis}\een  
so that we arrive at the important relation [see (\ref{onebis})]

\ben   \frac{\partial S_{ME}}{\partial \lambda_U} = (QG') \frac{\partial }{\partial \lambda_U} \int dV f[g(\lambda_N +\lambda_U U)] =\\=   (QG')   \int dV F'[g(\lambda_N + \lambda_U U)][g'(\lambda_N +\lambda_U U)] [\frac{\partial \lambda_N}{\partial \lambda_U}+U],                \label{five}\een that using (\ref{fourbis})  gives 

\ben   \frac{\partial S_{ME}}{\partial \lambda_U} =  \lambda_U \int dV U g'(\lambda_N +\lambda_U U) [\frac{\partial \lambda_N}{\partial \lambda_U}+U],   \label{fivebis}\een
which, according to  (\ref{three}) yields the Euler relation

\be \frac{\partial S_{ME}}{\partial \lambda_U} = \lambda_U \frac{\partial \langle U \rangle }{\partial \lambda_U } 
 \label{sixbis},      \ee so that 

\be  \frac{\partial S}{\partial <U>}= \frac{\partial S}{\partial \lambda_U} \frac{\partial \lambda_U}{\partial <U>}=  \lambda_U   \frac{\partial <U>}{\partial \lambda_U}  \frac{\partial \lambda_U}{\partial <U>}= \lambda_U,       \ee
the first reciprocity relation. Finally, introducing now the Jaynes’ parameter $\lambda_J$ (the Legendre
transform of $S_{ME}$)

\be  \lambda_J(\lambda_U) = S(<U>) - \lambda_U  <U>(\lambda_U),   \ee
it is clear that

 \be  \frac{\partial \lambda_J }{\partial \lambda_U}= \frac{\partial S_{ME}}{\partial <U>}\frac{\partial <U>}{\partial \lambda_U} - \lambda_U \frac{\partial <U>}{\partial \lambda_U} - <U> = - <U>,  \ee
the second reciprocity relation. There exists a thermodynamics associated to the general entropic forms under study here, since Jaynes' MaxEnt approach successfully works.

\section{Renyi's entropy $S_R$ and reciprocity relations}

\nd We specialize the preceding discussion for  $S_R$ and ascertain, as should be expected, that it does work. 
This fact notwithstanding, if we try to explicitly write $\xi_{ME}$, as we will do in the following Section, 
problems arise. Because of such contradiction, it is worthwhile to repeat the preceding argument for $S_R$. One  has 

\be S=Q \ln{[\int dV f(\xi)]},   \ee
that becomes Renyi's one for $Q=(1-q)^{-1}$ and $f(\xi)=\xi^q$. One abbreviates also $h= \int dV f(\xi)$.
 Then
\be S'= (Q/h) \int dV f'(\xi).\ee Define $F=(Q/h)f'(\xi) $ and consider the inverse function of $F$, namely,

\be g= F^{-1};\,\,\,FF^{-1}(\nu)= \nu.\ee 
In this case 

\be F= (Q/h)q\xi^{q-1};\,\,\,g(\xi)= [(h/Q)(\xi/q)]^{\frac {1} {q-1}};\,\,\,gF(\xi)= \xi,\ee
%%%%%%%%%\be \xi_{ME} = [(h/Q) (\lambda_N +\lambda_U U)/q]^{\frac {1} {q-1}}.\ee

The MaxEnt variational problem ends up being

\be F - \lambda_N -\lambda_U U=0,\ee so that the MaxEnt solution $\xi_{ME}$ is

\be  \xi_{ME}= g(\lambda_N +\lambda_U U), \label{one}\ee and the MaxEnt entropy reads

\be  S_{ME}= Q \ln{[\int dV f[g(\lambda_N +\lambda_U U)]]}.\label{two}\ee One also has

\be \frac{\partial \langle U \rangle }{\partial \lambda_U }= 
  \int dV U g'(\lambda_N +\lambda_U U) [\frac{\partial \lambda_N}{\partial \lambda_U}+U], \label{three}\ee  and

\ben 0= \frac{\partial }{\partial \lambda_U}\int dV \xi =\\= 
   \int dV g'(\lambda_N +\lambda_U U) [\frac{\partial \lambda_N}{\partial \lambda_U}+U]=0,  \label{four}\een  
so that we arrive at the important relation [see (\ref{one})], after remembering that $F'g(\nu)=\nu$,

\ben   \frac{\partial S_{ME}}{\partial \lambda_U} = (Q/h) \frac{\partial }{\partial \lambda_U} \int dV f[g(\lambda_N +\lambda_U U)] =\\=      \int dV F'[g(\lambda_N + \lambda_U U)][g'(\lambda_N +\lambda_U U)] [\frac{\partial \lambda_N}{\partial \lambda_U}+U],                \label{fiveone}\een that using (\ref{four})  gives 

\ben   \frac{\partial S_{ME}}{\partial \lambda_U} =  \lambda_U \int dV U g'(\lambda_N +\lambda_U U) [\frac{\partial \lambda_N}{\partial \lambda_U}+U],   \label{fivetwo}\een
which, according to  (\ref{three}) yields the Euler relation

\be \frac{\partial S_{ME}}{\partial \lambda_U} = \lambda_U \frac{\partial \langle U \rangle }{\partial \lambda_U } 
 \label{six},      \ee so that 

\be  \frac{\partial S}{\partial <U>}= \frac{\partial S}{\partial \lambda_U} \frac{\partial \lambda_U}{\partial <U>}=  \lambda_U   \frac{\partial <U>}{\partial \lambda_U}  \frac{\partial \lambda_U}{\partial <U>}= \lambda_U,       \ee
the first reciprocity relation. Finally, introducing now the Jaynes’ parameter $\lambda_J$ (the Legendre
transform of $S_{ME}$)

\be  \lambda_J(\lambda_U) = S(<U>) - \lambda_U  <U>(\lambda_U),   \ee
it is clear that

 \be  \frac{\partial \lambda_J }{\partial \lambda_U}= \frac{\partial S_{ME}}{\partial <U>}\frac{\partial <U>}{\partial \lambda_U} - \lambda_U \frac{\partial <U>}{\partial \lambda_U} - <U> = - <U>,  \ee
the second reciprocity relation.

\section{MaxEnt-Renyi's peculiarities}

\nd This was detected in \cite{11}. Let us return to 

\be h = \int dV \xi^q, \ee

\ben F=(Q/h)f'(\xi);\\ F - \lambda_N -\lambda_U U=0,   \label{hidden}\een 
and   set $f(\xi)= \xi^q$ (specify Renyi's case, with $Q=1/(1-q)$). One has  

\be  \frac{q\xi^{q-1}}{(1-q)h} - \lambda_N -\lambda_U U=0, \ee
and now integrate  (\ref{hidden}) over $\xi dV$. {\it The $h$ in the denominator cancels the 
$\xi-$integral in the numerator!} Thus, 

\ben 0= \int  dV \xi \{\frac{(1-q)^{-1}q \xi^{q-1}}{h} - \lambda_N -\lambda_U U\}=\\=  
\frac{q}{1-q} -\lambda_N -\lambda_U \langle U \rangle =0, \label{varia}\een leading to

\be  \lambda_N = \frac{q}{(1-q)} - \lambda_U  \langle U \rangle,
  \label{landaN}       \ee   which {\it diverges} for $q==1$. Also, one has
		
				%%%%%%%%%%%%%%\subsection{The Renyi failure}
				
				\be \frac{\partial \lambda_N  }{\partial  \lambda_U} = - \langle U \rangle 
				-\lambda_U \frac{\partial   \langle U \rangle }{\partial  \lambda_U},    \label{nova} \ee
	which is an interesting Renyi relation, to be compared to the Legendre transform (see Sect.1)
		
		\be  \lambda_J = S- \lambda_U  \langle U \rangle, \ee so that 
		
		\be \frac{\partial \lambda_J  }{\partial  \lambda_U} = -  \langle U \rangle.   \ee
		
		\nd Summing up, we have a  divergence at $q=1$ in (\ref{landaN}) that we should try to understand.
		In order to do so, we embark now on a detour.

\section{Comparison with Tsallis' case}

\nd It is of interest to compare (\ref{landaN}) above with its Tsallis' counterpart $S_T$. One has \cite{tsallis,web,ugur,tempesta}

\be S_T = \frac{1}{q-1}  \int dV [\xi  - \xi^q].  \label{t1}\ee 
Thus, 

\be    \int dV \xi \frac{\partial S_T}{\partial \xi  }=  \frac{1}{q-1}  \int dV [\xi  - q\xi^q],
 \label{t2} \ee
leading to 

\be dV \frac{1}{q-1}  \int dV [\xi  - q\xi^q] - \lambda_N - \lambda_U  \langle U \rangle=0. \label{t3}\ee
Now, using $\xi$-normalization, this can be recast, using $\int dV \xi^q= 1 + (1-q) S_T,$
 
\be qS_T - 1 -   \lambda_N - \lambda_U  \langle U \rangle=0,   \label{t4}\ee that should be compared to 
(\ref{lambdaN}). Now, after adding and substracting $S_T$ on the l.h.s.,

\be  \lambda_J= S_T -   \lambda_U  \langle U \rangle=  \lambda_N +1 + (1-q)  S_T.    \label{t5}\ee
In the limit $q \rightarrow 1$ this yields

\be \lambda_J= \lambda_N +1, \label{t6}\ee
the Boltzamnn-Gibbs classical result \cite{jaynes}.

\subsection{Tsallis' case alternative viewpoint}

\nd This alternative situation is also mentioned in \cite{rg}. Quite simply, $S_T$ can cast in two identical  fashions:

\be S_T = \frac{1}{q-1}  \int dV [\xi  - \xi^q]=  \frac{1}{q-1} - 
 \frac{1}{q-1}  \int dV  \xi^q,   \label{t6}\ee which gives rise to two different variational problems. Using the second form one has

\be    \int dV \xi \frac{\partial S_T}{\partial \xi  }=  \frac{q}{1-q}  \int dV \xi^q, 
 \label{t8} \ee and

\be     \frac{q}{1-q}  \int dV \xi^q   - \lambda_N - \lambda_U  \langle U \rangle=0.   , 
 \label{t7} \ee Using again $\int dV \xi^q= 1 + (1-q) S_T,$ we are led to, after adding and substracting %S_T$,

\be  \lambda_J= S_T -   \lambda_U  \langle U \rangle=  \lambda_N -  \frac{q}{1-q}  + 
 (1-q)  S_T,     \label{t9}\ee which in the limit $q \rightarrow 1$ results in a divergence!\vskip 3mm 

\nd What is happening here? The answer is that casting $S_R$ as

\be S_T =   \frac{1}{q-1} - 
 \frac{1}{q-1}  \int dV  \xi^q,   \label{t10}\ee
assumes normalization from the very beginning, which {\it unduly restricts} the $\xi$-variational space. Additionally, we keep $\lambda_N$ in the variation, which is an inconsistency. Thus, the accompanying divergence.

\subsection{Surrogate Renyi entropy $S_R^S$}
We just add zero to the Renyi definition in the fashion ($b$ is a constant to be chosen later on) 

\be  S_R^S= \frac{bq}{1-q} \ln{\left[\int dV \xi\right]} + \frac{1}{1-q} \ln{\left[\int dV \xi^q\right]},    \label{t11} \ee
and thus

\be \int dV \xi \frac{\partial S_R^S}{\partial \xi} =  \frac{q}{1-q} (b +1),\ee which vanishes if one 
 chooses $b=-1$. Accordingly,

\be \lambda_N= -\lambda_U    \langle U \rangle, \ee and the divergence has disappeared. We have discovered then that the original Renyi definition somehow assumes normalization from the beginning. In any case, $\lambda_N$ does not tend to its Shannon counterpart as $q$ tends to unity.

\section{Conclusions}

\nd We have investigated here quite general entropies that lack trace form, a family of entropic functionals that includes as a distinguished member the celebrated Renyi entropy $S_R$.   We
 studied for such family the validity of thermodynamics'  reciprocity relations (RR) that would, in turn,  legitimate   an associated statistical mechanics, complying with the basic thermodynamics' tenets. We proved the RR exist  for all possible entropies.

\vskip 2mm

\nd Our endeavors both illuminated and   allowed us to understand the origin of the MaxEnt-Renyi peculiarities, discovered  in 
\cite{11}, that seemingly impaired the associated RR for the particular entropic form $S_R$. Amongst these peculiarities, we single out a singularity in Eq. (\ref{landaN}), connecting $\lambda_N$ (normalization multiplier) with $\lambda_U$ (energy multiplier) that emerges in the limit $q \rightarrow 1$. We found that the singularity can be removed if one 
 replaces  $S_R$ by a surrogate quantifier  $S_R^S$, essentialy equivalent to  $S_R$ but not identical to it. These two measures are given by different functionals  of the probability density $\xi$, whose numerical values coincide if one explicitly takes into account the normalization of $\xi$.

\vskip 3mm  \nd 
A similar artifact is seen to apply to Tsallis entropy $S_T$ when extremized with linear constraints. $S_T$ can be cast in two different ways as a function 
of the probability density $\xi$, and for one of them a similar singularity emerges as well. In the alternative instance, instead, the $q \rightarrow 1$ limit makes Tsallis normalization multiplier to converge to the Shannon one.

\vskip 3mm  \nd We conclude that, if for some entropic form $S_A$ the MaxEnt treatment displays a singularity, there should be an alternative way of writing $S_A$ that overcomes the difficulty, as we have proved in Section 3  that RR are valid for any $S_A$.


\newpage

\begin{thebibliography}{99}

\bibitem{1} C. M. Herdman, Stephen Inglis, P.-N. Roy, R. G. Melko, and A. Del
Maestro, Phys. Rev. E {\bf 90}, 013308 (2014).

\bibitem{2} Mohammad H. Ansari and Yuli V. Nazarov, Phys. Rev. B {\bf 91}, 174307
(2015).

\bibitem{3} Lei Wang and Matthias Troyer,      Phys. Rev. Lett.  {\bf 113},  110401
(2014).

\bibitem{4}  Matthew B. Hastings, Iv�n Gonz�lez, Ann B. Kallin, and Roger G. Melko, Phys. Rev. Lett {\bf 104},  157201  (2010).

\bibitem{5}  Richard Berkovits, Phys. Rev. Lett.  {\bf 115}, 206401
(2015).

\bibitem{6} Nima Lashkari,
   Phys. Rev. Lett. {\bf 113}, 051602   (2014).


\bibitem{7}  Gabor B. Halasz and Alioscia Hamma,
   Phys. Rev. Lett. {\bf 110},   170605    (2013).

\bibitem{8}  MB Hastings, I Gonz�lez, AB Kallin, RG Melko,
   Phys. Rev. Lett. {\bf 104},  157201     (2010);
   A. De Gregorio, S.M. lacus, {\bf 179}, 279 (2009).


   \bibitem{9} Leila Golshani, Einollah Pasha, Gholamhossein Yari,
   Information Sciences, {\bf 179},  2426 (2009); J.F. Bercher, Information Sciences {\bf 178},  2489 (2008).


\bibitem{10}   EK Lenzi, RS Mendes, LR da Silva,
 Physica A {\bf 280}, 337 (2000).

\bibitem{zentro}  P. Tempesta, Proc. R. Soc. A {\bf 472} (2016) 20160143.
	
	\bibitem{jaynes} E.T. Jaynes, in: {\it Statistical physics}, ed. by W.K. Ford (Benjamin,
New York, 1963);
A. Katz, {\it Statistical mechanics} (Freeman, San Francisco,
1967).

\bibitem{11} A. Plastino, M. C. Rocca, F. Pennini, Phys. Rev. E  {\bf 94} (2016) 012145.

%\bibitem{tsallis}  M. Gell-Mann and C. Tsallis, Eds. {\it Nonextensive Entropy:
%Interdisciplinary applications}, Oxford University Press, Oxford,
%2004;  C. Tsallis, {\it Introduction to Nonextensive Statistical
%Mechanics: Approaching a Complex World}, Springer, New York, 2009.

%\bibitem{web} See http://tsallis.cat.cbpf.br/biblio.htm for a
%regularly updated bibliography on the subject.

\bibitem{book} R. B. Lindsay, H. Margenau, {\it Foundations of physics} (Dover, NY, 1957).

\bibitem{deslogue} E. A. Desloge, {\it Thermal physics} NY, Holt, Rhinehart and
   Winston, 1968.


\bibitem{universal} A. Plastino, A. R. Plastino, Phys. Lett. A {\bf 226} (1997) 257.


\bibitem{tsallis}  M. Gell-Mann and C. Tsallis, Eds. {\it Nonextensive Entropy:
Interdisciplinary applications}, Oxford University Press, Oxford,
2004;  C. Tsallis, {\it Introduction to Nonextensive Statistical
Mechanics: Approaching a Complex World}, Springer, New York, 2009.

\bibitem{web} See http://tsallis.cat.cbpf.br/biblio.htm for a
regularly updated bibliography on the subject.

\bibitem{ugur} U. Tirnakli, E. P. Borges, Scientific Reports {\bf 6}, Article number: 23644 (2016).

\bibitem{tempesta} G. Sicuro, P. Tempesta, A. Rodríguez, C. Tsallis, Annals of Physics {\bf 363} (2015) 316.

\bibitem{rg} See Research Gate DOI10.13140/RG.2.2.26577.02403.


%\bibitem{abe} Sumiyoshi Abe, S. Martinez, F. Pennini,
% A. Plastino, Phy. Lett. A {\bf 281} (2001) 126.

%%\bibitem{evaldo} E. M. F. Curado, C. Tsallis, J. Phys. A {\bf 24} (1991) L69.

%%\bibitem{mendes} C. Tsallis, R. S. Mendes, A.R. Plastino, Physica A {\bf 261} (1998) 534.

%%\bibitem{secondvar} A. Plastino, M. C. Rocca, Physica A {\bf 436} (2015) 572.
\end{thebibliography}
\end{document}